\documentclass[prx, twocolumn, superscriptaddress, reprint, aps, floatfix]{revtex4-1}
\usepackage{empheq}
\usepackage{amsmath} 
\usepackage{amsfonts}
\usepackage{graphicx}
\usepackage{float}
\usepackage{cleveref}
\usepackage{bm}
\usepackage{color}
\usepackage[colorinlistoftodos]{todonotes}
\newcommand{\toggl}[1]{}

\begin{document}

\title{The free surface of a colloidal chiral fluid: waves and instabilities from odd stress and Hall viscosity}

\author{Vishal Soni*}
\affiliation{James Franck Institute  and Department of Physics, University of Chicago, Chicago, IL 60637, USA}
\author{Ephraim Bililign*}
\affiliation{James Franck Institute  and Department of Physics, University of Chicago, Chicago, IL 60637, USA}
\author{Sofia Magkiriadou*}
\affiliation{James Franck Institute  and Department of Physics, University of Chicago, Chicago, IL 60637, USA}
\author{Stefano Sacanna}
\affiliation{Molecular Design Institute, Department of Chemistry, New York University, 29 Washington Place, New York, New York 10003, USA}
\author{Denis Bartolo}
\affiliation{University of Lyon, ENS de Lyon, University of Claude Bernard, CNRS, Laboratoire de Physique, F-69342 Lyon, France}
\author{Michael J. Shelley}
\affiliation{Center for Computational Biology, Flatiron Institute, New York, NY
10010} 
\affiliation{Courant Institute, NYU, New York, NY 10012}
\author{William T. M. Irvine}
\email{wtmirvine@uchicago.edu \\ *These authors should be seen as first authors on equal footing}
\affiliation{James Franck Institute, Enrico Fermi Institute and Department of Physics, University of Chicago, Chicago, IL 60637, USA}

\begin{abstract}
In simple fluids, such as water, invariance under parity and time-reversal symmetry imposes that the rotation of constituent ``atoms'' are determined by the flow and that viscous stresses damp motion. 
Activation of the rotational degrees of freedom of a fluid by spinning its atomic building blocks breaks these constraints and has thus been the subject of fundamental theoretical interest across classical and quantum fluids~\cite{lenz_membranes_2003, yeo_rheology_2010,furthauer_active_2013, nguyen_emergent_2014, goto_purely_2015, yeo_collective_2015, climent_dynamic_2006,avron_viscosity_1995, wiegmann_anomalous_2014,ariman_microcontinuum_1973,rosensweig2013,Rinaldi2014}. However, the creation of a model liquid which isolates chiral hydrodynamic phenomena has remained experimentally elusive.
Here we report the creation of a cohesive two-dimensional chiral liquid consisting of millions of spinning colloidal magnets and  study  its flows. 
We find that dissipative viscous “edge pumping” is a key and general mechanism of chiral hydrodynamics, driving uni-directional surface waves and instabilities, with no counterpart in conventional fluids. 
Spectral measurements of the chiral surface dynamics reveal the presence of Hall viscosity, an experimentally long sought property of chiral fluids~\cite{avron_viscosity_1995,avron_odd_1998,abanov2018,banerjee_odd_2017}.
Precise measurements and comparison with theory demonstrate excellent agreement with a minimal but complete chiral hydrodynamic model, paving the way for the exploration of chiral hydrodynamics in experiment. 
\end{abstract}
\maketitle

Hydrodynamic theories describe the flow of systems as diverse as water, quantum electronic states~\cite{bandurin_negative_2016}, and galaxies~\cite{pringle_astrophysical_2007} over decades in scale~\cite{secchi_massive_2016}.
Since hydrodynamic equations are built on symmetry principles and conservation laws alone, systems with similar symmetries have similar descriptions and flow in the same way. 

For example, symmetry under parity and time reversal -- conditions  met by all conventional fluids at thermal equilibrium --
constrains both the stress and viscosity tensors to be symmetric. 
These constraints are in principle alleviated in collections of interacting units that are driven to rotate~\cite{tsai_chiral_2005,scaffidi_hydrodynamic_2017,wiegmann_anomalous_2014,banerjee_odd_2017,van_zuiden_spatiotemporal_2016,furthauer_active_2013,avron_viscosity_1995,snezhko_complex_2016,kokot2017}.
This seemingly innocent twist on an otherwise structureless fluid represents, however,  an elemental change with rich hydrodynamic consequences common to quantum Hall fluids, vortex fluids, and chiral condensed matter.
Collections of spinning particles offer a natural opportunity to engineer and study the properties of such chiral fluids; 
experimental examples include rotating bacteria~\cite{petroff_fast-moving_2015}, colloidal and millimeter-scale magnets~\cite{grzybowski_dynamic_2000,grzybowski_dynamic_2001,grzybowski_dynamics_2002,grzybowski_dynamic_2002,belovs_hydrodynamics_2016,yan_rotating_2014,yan_jing_colloidal_2015}, ferrofluids in rotating magnetic fields~\cite{rosensweig2013,Rinaldi2014}, and shaken chiral grains~\cite{tsai_chiral_2005,scholz_rotating_2018}. 
Such systems have been shown to have non-trivial dynamics. For example, ferrofluids driven by AC fields can flow against external pressure~\cite{Bacri1995} and small numbers of spinning particles  self-assemble into dynamic crystalline clusters~\cite{grzybowski_dynamic_2000,grzybowski_dynamic_2001,grzybowski_dynamics_2002,grzybowski_dynamic_2002, climent_dynamic_2006, yan_rotating_2014,yan_jing_colloidal_2015}.

\begin{figure*}[htb]
\centering
\includegraphics[width=\textwidth]{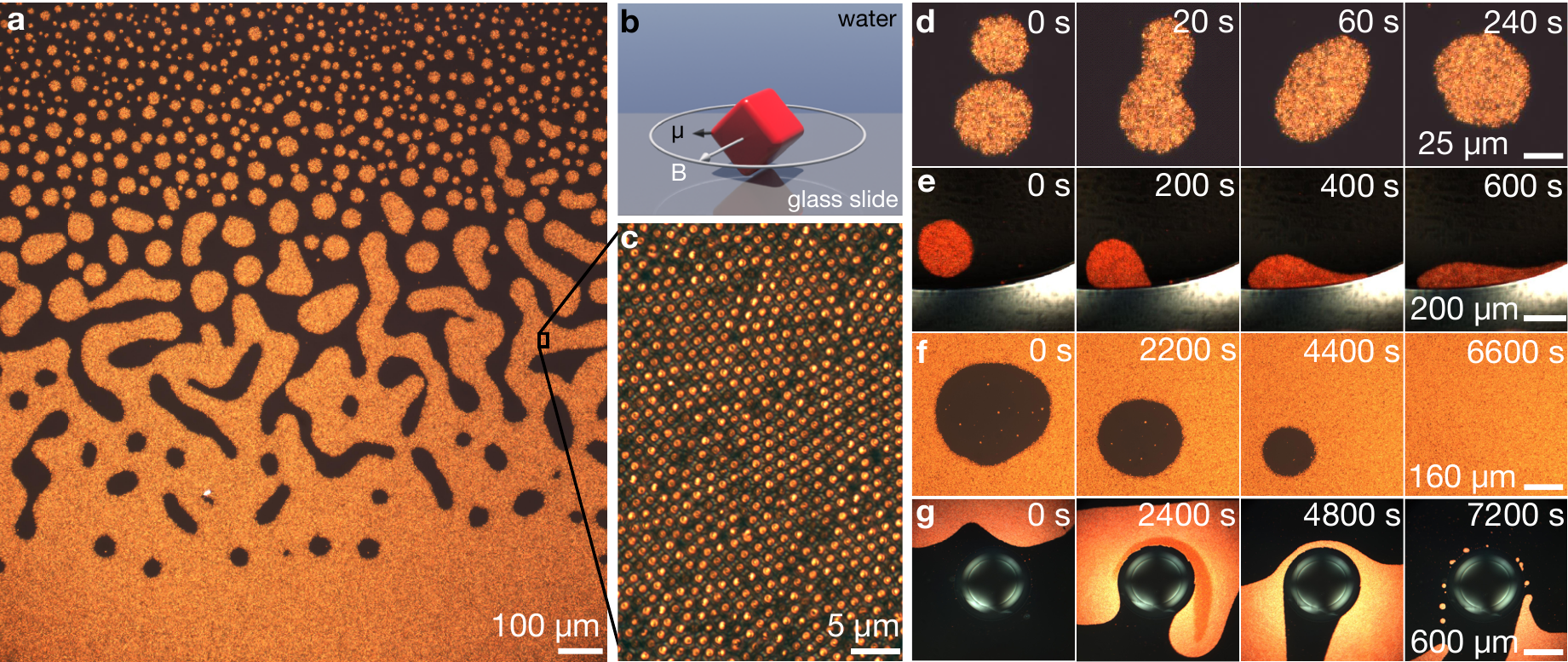}
\caption{\label{fig:chiralfluid} {\bf A chiral fluid of spinning  colloidal magnets.} {\bf a,} Optical micrograph of the colloidal magnets in bulk, after a few minutes of spinning. {\bf b,} Schematic of one colloidal particle. The $\sim$1.6 $\mu m$ hematite colloidal cubes have a permanent magnetic moment ($\mu$, black arrow). They are suspended in water, sedimented onto a glass slide, and spun by a rotating magnetic field ($B$, white arrow tracing the white circle). 
{\bf c,} Optical micrograph of the bulk colloidal magnets at increased magnification.
{\bf d-g,} The particles attract and form a cohesive material with an apparent surface tension that, over timescales from minutes to hours, behaves like a fluid: {\bf d}, clusters coalesce and {\bf e} spread like liquid droplets when sedimented against a hard wall; {\bf f} void bubbles collapse; and {\bf g} when driven past an obstacle, the fluid flows around it, thinning and eventually revealing an instability to droplet formation. All images were taken through crossed polarizers.}
\end{figure*}

\section{A colloidal chiral fluid}
We report the creation of a millimeter-scale cohesive chiral fluid (Fig.~\ref{fig:chiralfluid}a) by spinning millions of colloidal magnets with a magnetic field (Figs.~\ref{fig:chiralfluid}b,~\ref{fig:chiralfluid}c), and we track its flows over hours (see Supplementary Movies 1, 2). 
The macroscopic flow of our chiral fluid is reminiscent of free surface flows of Newtonian fluids: 
nearby droplets merge (Fig.~\ref{fig:chiralfluid}d and Supplementary Movie 3), fluid spreads on a surface under the influence of gravity (Fig.~\ref{fig:chiralfluid}e and Supplementary Movie 4), voids collapse (Fig.~\ref{fig:chiralfluid}f and Supplementary Movie 5), and thin streams go unstable, as revealed by flowing fluid past a solid object (Fig.~\ref{fig:chiralfluid}g and Supplementary Movie 6). We demonstrate  that these seemingly familiar features are accompanied by unique free surface flows. We then exploit the odd interfacial dynamics of this prototypical chiral liquid to infer its material constants, which remain out of reach of conventional rheology.  

In contrast to Newtonian fluids,  the surface of our fluid supports a spontaneous unidirectional edge flow in its rest state, as well as unusual morphological dynamics such as the rotation of asymmetric droplets illustrated in Supplementary Movie 3. 

\section{Chiral surface waves and `edge pumping'}
\begin{figure*}
\centering
\includegraphics[width=1\textwidth]{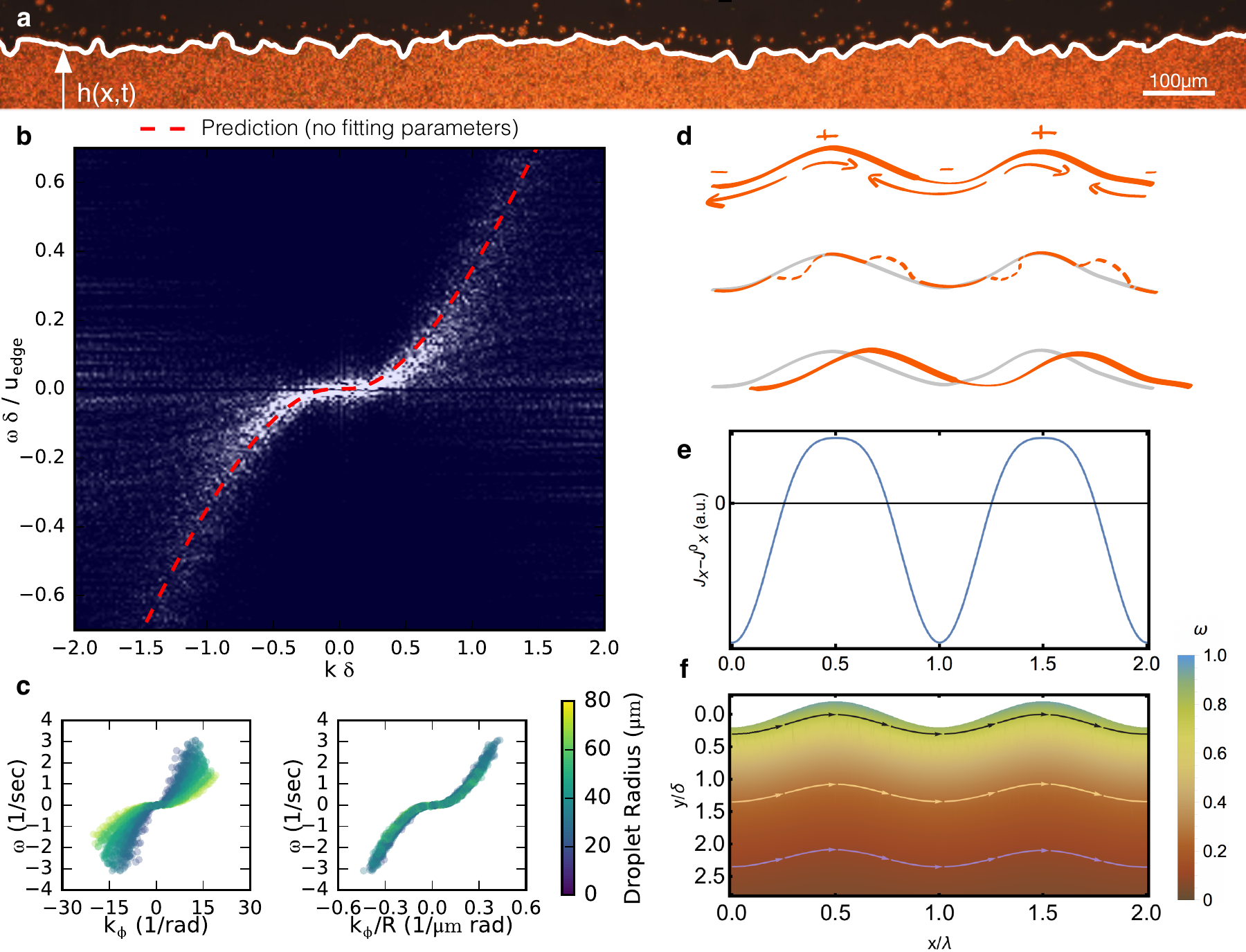}
\caption{\label{fig:fig3} {\bf Surface waves in a chiral spinner fluid.} {\bf a,} Surface waves are excited by perturbing a strip of the spinner fluid. 
To characterize them, we track the height profile of the strip in time, $h(x,t)$. 
{\bf b,} The resulting power spectrum from these waves $\langle|h(k,\omega)|\rangle$ is plotted versus the normalized wave vector $k\delta$ and frequency $\omega/(u_{\rm edge}/\delta)$. The spectrum is peaked on a curve corresponding to the dispersion relation of the waves. Shown with the red dashed line is the theoretical prediction for the dispersion relation, obtained with the hydrodynamic parameters that we measure in Fig.~\ref{fig:droplets}; its long-wavelength asymptotic form is given in Eq.~\eqref{Eq:dispersion}. 
{\bf c,} The power spectrum $\omega(k)$ for surface waves on a perturbed circular droplet of spinner fluid (left panel, see Supplementary Information) can be collapsed (right panel) by rescaling the angular wavenumber $k$ by the droplet radius $R$.
{\bf d,} Sketch of the mechanism for wave propagation. The propagation of waves can be understood by considering the mass flux, plotted in {\bf e}. The chiral fluid is displaced from the high curvature to the low curvature regions. This process explicitly breaks the left-right symmetry, thereby propagating surface waves along only one direction. 
{\bf e,} Correction to the net mass flux along the interface due to a sinusoidal height perturbation, $J_{x} - J_{x}^{o}$, where $J_{x}^{o}$ is the mass flux in a flat strip and $J_{x}$ is the mass flux in the presence of a perturbation. {\bf f,} Predicted vorticity field from the chiral-hydrodynamic model (see Supplementary Information). 
}
\label{fig:spectrum}
\end{figure*}

To investigate these lively surface flows, we first look at surface excitations in a simple slab geometry, as shown in Fig.~\ref{fig:spectrum}a and Supplementary Movie 7.
We measure the spectrum of surface fluctuations, $|h(k,\omega)|^2$, by tracing the height profile, $h(x,t)$, of the surface and Fourier-transforming it in space and time. 
We observe the spectrum to be peaked along a curve $\omega(k)$, revealing the existence of dispersive waves (see Fig.~\ref{fig:spectrum}b). 
The curve has only one branch with odd parity, meaning that the waves are unidirectional. 
This behavior contrasts that of
conventional surface waves that propagate in all directions. 

\begin{figure}
 \centering\includegraphics[width=.5\textwidth]{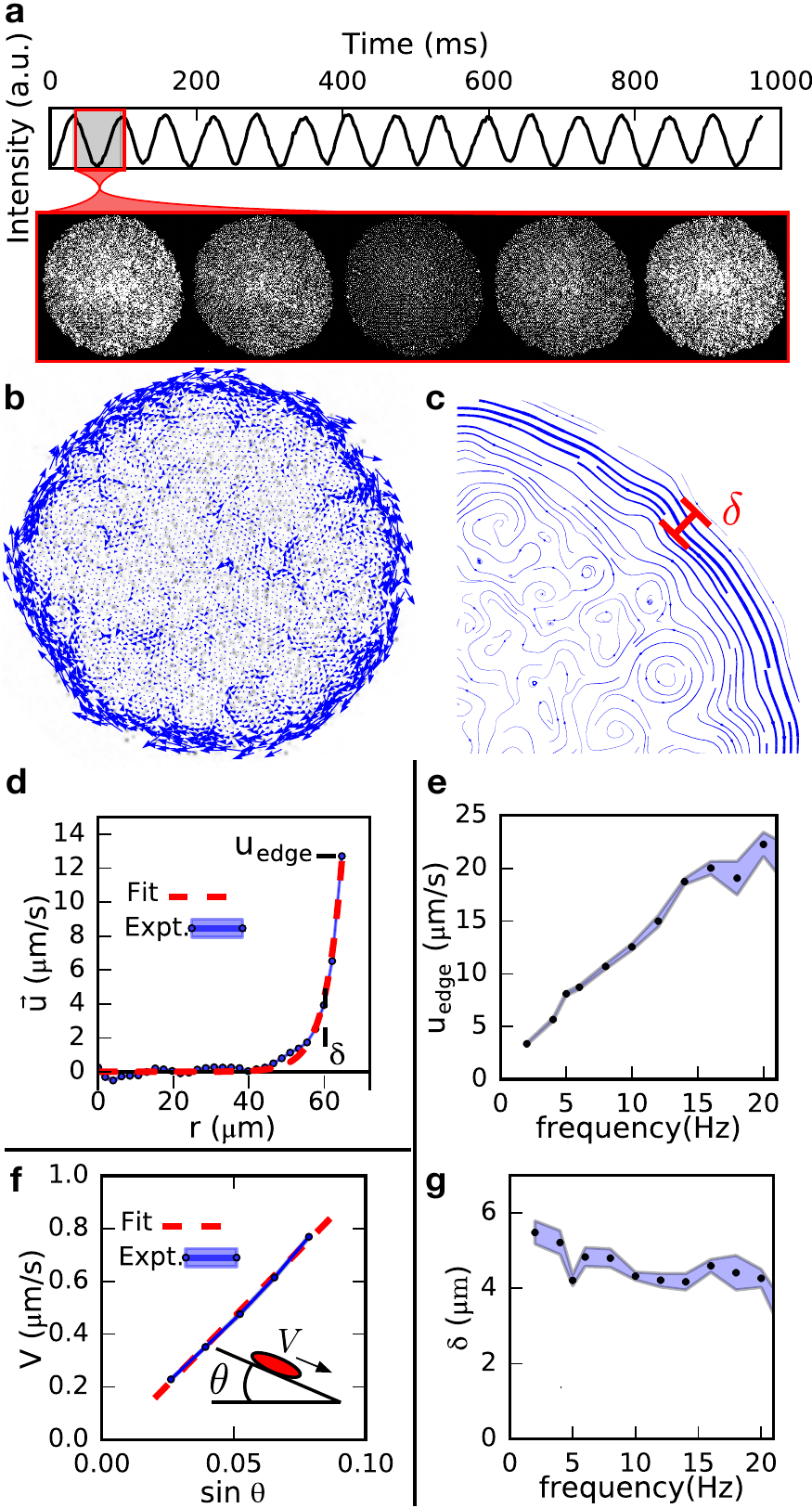}
\caption{{\bf Characterization of a droplet of chiral spinner fluid.} {\bf a,} When viewed through crossed polarizers, the particles blink as they spin. This allows us to confirm that they all spin at the same frequency, set by the rotating magnetic field. 
{\bf b,} By measuring the velocity of each particle within a cluster, we find a flow profile that is concentrated at the edge within a penetration layer $\delta$ shown in {\bf c, d,} and {\bf g}. {\bf c,} A zoomed-in view of the flow streamlines, obtained by averaging several instantaneous velocity profiles such as the one shown in {\bf b}.  
{\bf d,} By measuring the flow profile, the edge current $\mathrm{u}_{\rm edge}$ and penetration depth $\delta$ are extracted. {\bf e, g,} By measuring the flow profile $u(r)$ at a range of frequencies, we extract the shear viscosity, $\eta$, and rotational viscosity, $\eta_{R}$, in terms of the substrate friction, $\Gamma_{u}$. 
{\bf f,} Finally, by tilting a sample and measuring the sedimentation velocity of a droplet, we extract the substrate friction. 
}
\label{fig:droplets}
\end{figure}

These surface waves beg a hydrodynamic description. 
Chiral-fluid hydrodynamics  follows from conservation of momentum and angular momentum, and thus includes both the spinning rate of individual fluid particles as well as the momentum and angular momentum of their flow~\cite{furthauer_active_2013,bonthuis_electrohydraulic_2009,tsai_chiral_2005,dahler_theory_1963,huang_continuum_2010}.
Because our colloids are birefringent, we are able to measure their individual spinning rate by imaging through crossed polarizers. We find that all  particles rotate at the same rate, $\Omega$, which is set by the rotating magnetic field (see Fig.~\ref{fig:droplets}a and Supplementary Movie 8). 
From this it follows that the particles' rotational inertia is negligible; the torque exerted on each particle by the magnetic field instantly adjusts to balance the frictional torques exerted by the neighboring particles and the solid substrate. 
This fast response enables the decoupling of the angular momentum equation from the momentum equation.  Nonetheless a strong signature of the microscopic angular momentum manifests as an `odd' stress. 
A minimal hydrodynamic theory then balances the force generated by viscous and odd hydrodynamic stresses, $\partial_j \sigma_{ij}$, 
against friction with the substrate, $\Gamma_{u} v_i$, and surface tension  $\gamma$ at the fluid interface. 
In this theory, which has been used to capture the bulk flows of chiral granular fluids, the hydrodynamic stress tensor is given by:
\begin{align}
\sigma_{ij} &= -p \delta_{ij} + \eta \left(\partial_i v_j+\partial_j v_i\right) + \eta_{\rm R} \epsilon_{ij}\left(2\Omega-\omega\right).
\end{align}
$\sigma_{ij}$ includes the pressure $p$ and ordinary viscous stress also present in Newtonian fluids with a shear viscosity $\eta$. 
The additional term containing the Levi-Civita symbol $\epsilon_{ij}$ and the rotational viscosity $\eta_R$, captures the rotational friction between neighboring particles~\cite{dahler_theory_1963,bonthuis_electrohydraulic_2009,tsai_chiral_2005,furthauer_active_2013}. 
Such an odd stress builds up as the local spinning rate $\Omega$ deviates from half the local fluid vorticity $\omega = \hat{z}\cdot (\nabla \times v)$. 
In torque-free fluids, angular momentum conservation constrains these two quantities to be equal: odd stresses are unique to chiral fluids.

We note that there is no direct appearance of the magnetic field or its stresses in this hydrodynamic description unlike in conventional ferrofluids.  In this respect, our colloidal chiral fluid can be seen as a special type of driven ferrofluid in which the only role of magnetic forces is to induce chirality.

To make a quantitative comparison between our model and the flows we observe, we require a measurement of the  hydrodynamic and friction coefficients $\eta$, $\eta_{\rm R}$, and $\Gamma_{u}$.  
Fortunately, the prominent effect of odd stress at the free surface of our chiral fluid can be effectively exploited to infer its bulk rheology.
The homogeneous spinning motion of the colloidal particles gives rise to a net tangential edge flow even in the absence of pressure gradients. 
These tread-milling dynamics, characteristic of all chiral fluids~\cite{tsai_chiral_2005,nguyen_emergent_2014,petroff_fast-moving_2015,yan_rotating_2014,van_zuiden_spatiotemporal_2016}, are illustrated in circular droplets in Figs.~\ref{fig:droplets}b-e and Supplementary Movie 9. The tangential flow that is localized at the free surface is readily explained by expressing the hydrodynamic equation in terms of vorticity for an incompressible chiral fluid:
\begin{equation}
\left(\nabla^2-{\delta^{-2}}\right)\omega=0
\label{Eq:vorticity}
\end{equation}
where $\delta=\sqrt{{(\eta+\eta_{R})}/{\Gamma_{u}}}$. This Helmholtz 
equation indicates that the vorticity generated at the surface decays exponentially into the fluid, with a characteristic penetration depth $\delta$ (see Figs.~\ref{fig:droplets}c, d, g).  
In this model, the absence of substrate friction causes the penetration depth to diverge, resulting in rigid-body rotation of the entire fluid, as observed in ferrofluid droplets~\cite{Bacri1994}. 
The magnitude of the vorticity at the free surface, $\omega_{\rm edge}=2\Omega\,\eta_{\rm R}/(\eta+\eta_{\rm R})$, is set by  the stress-free boundary condition for a flat strip and expresses the competition between the  odd and viscous stresses (see Supplementary Information). 
We point out that $\omega_{\rm edge}$ is directly proportional to  $\eta_R$, which demonstrates the importance of odd stress for the dynamics.
Comparison between experiment and prediction (Fig.~\ref{fig:droplets}d) yields the values of $\eta$ and $\eta_{R}$ in terms of  $\Gamma_{u}$. The latter is then  measured by tilting the substrate and measuring the sedimentation rate of droplets (see Fig.~\ref{fig:droplets}f, and  Supplementary Information).
Ultimately, we find $\eta= 4.9\pm 0.2 \times 10^{-8}\ \rm Pa\ m\ s$, $\eta_{R} = 9.1\pm 0.1\times 10^{-10}\ \rm Pa\ m\ s$, and $\Gamma_{u}=2.49\pm 0.03 \times 10^{3}\rm\ Pa\ s/m$.

Equipped with the hydrodynamic coefficients we can now investigate the origin of the surface waves within our model.  The mass flux in the tangential surface flow provides significant insight. 
This flow, sketched in Fig.~\ref{fig:spectrum}d and plotted in Figs.~\ref{fig:spectrum}e-f,
is determined by the balance of the tangential odd stress at the boundary, the shear stress, and the substrate friction. 
In the presence of a perturbation that varies the curvature of the interface, resistance to flow due to the shear stress will be modulated. 
For a sinusoidal perturbation, there is  enhanced flow in positively curved regions (top of the wave) and decreased flow in negatively curved regions (bottom of the wave). 
This `edge-pumping' moves material away from curved regions towards the flat wave front, giving rise to uni-directional wave motion. 

A linear stability analysis of the hydrodynamic equations (see Supplementary Information for a detailed calculation) confirms this scenario and yields a prediction for the dispersion relation, dissipation rate, and flow fields of surface waves, which we plot in Fig.~\ref{fig:spectrum}b (red dashed curves). 
With no fitting parameters, our model shows excellent agreement with the experimentally measured dispersion relation.  
For surface waves $h\sim e^{i(k x + \omega t)} $
of long wavelength $k \ll 1/\delta$, the asymptotic dispersion relation is:
\begin{equation}
 \omega(k)
= 2 \omega_{\rm edge} \frac{\eta_{\rm R}}{\eta + \eta_{\rm R}} (k \delta)^3 = 2  u_{\rm edge}  \frac{\eta}{\Gamma_{u}} k^3. 
\label{Eq:dispersion}
\end{equation}
where $u_{\rm edge}=2\frac{\eta_R}{\eta+\eta_R}\Omega \delta$.

The wave dynamics are thus crucially sensitive to  boundary layer flows.
A natural avenue for investigation, then, is to seek to increase the thickness of the boundary in order to increase its relative role. 
We now show how a slight increase of the penetration depth of the boundary layer amplifies chiral effects and reveals a long sought-after source of stress, commonly referred to as Hall viscosity. 

\section{Chiral wave damping and measurement of Hall Viscosity}
\begin{figure*}[htb]
 \centering\includegraphics[width=1\textwidth]{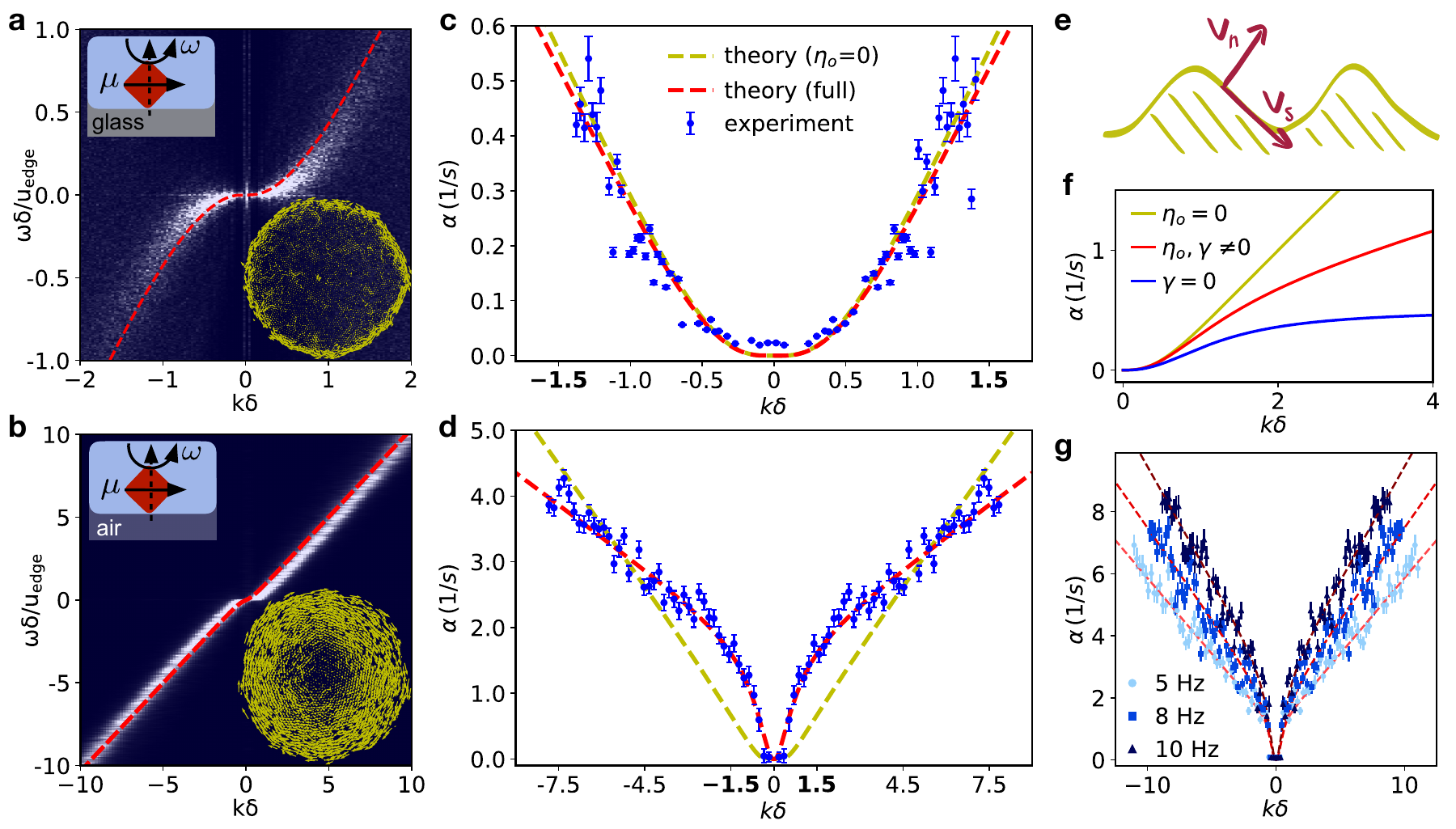}
\caption{{\bf Wave dissipation and measurement of Hall viscosity.} {\bf a,} In the circular geometry, surface waves yield power spectra $\langle |R(k,\omega)|\rangle$, plotted here versus the normalized wave vector $k\delta$ and frequency $\omega\delta/u_\mathrm{edge}$ (c.f. Fig.~\ref{fig:spectrum}c for a collection of spectra).
{\bf b,} Power spectrum at a low friction air-water interface, for which the edge current is delocalized into the bulk when compared to a high friction interface as in {\bf a} (see insets).
{\bf c,} The dissipation rate of waves on the surface of a circular droplet can be used to extract the surface tension, and the shape of $\alpha(k)$ can be captured by a theory with no odd viscosity ($\eta_{\rm o}=0$).
{\bf d,} Lowering substrate friction causes the dissipation to level-off for large $k\delta$, which can only be captured by a theory including $\eta_{\rm o}$.
{\bf e,} The tangential and normal components of velocity at the boundary give rise to a normal Hall stress (Eq.~\eqref{Eq:Hallstress}).
{\bf f,} The dissipation for a chiral fluid with $\eta_{\rm o}$ in the absence of surface tension, $\gamma$, vs. the same for a fluid with $\gamma$ in the absence of $\eta_{\rm o}$. For small $k\delta$, the two curves are indistinguishable. For large $k\delta$, the $\eta_{\rm o}$-dissipated fluid shows no $k$-dependence, while the $\gamma$-dissipated fluid shows linear $k$-dependence. Shown also for reference is the attenuation for a fluid with finite values of both $\gamma$ and $\eta_{\rm o}$
{\bf g,} With $\eta_{\rm o}$-induced attenuation, $\alpha(k)$ varies with frequency with all other parameters held constant, a trend that is not observed for $\gamma$-dissipated fluids.
}
\label{fig:etao}
\end{figure*}

We reduce the surface friction by allowing our chiral liquid to sediment upon an air-water interface (Fig.~\ref{fig:etao}b), as opposed to a glass surface (Fig.~\ref{fig:etao}a).
Due to the difficulty in maintaining a slab geometry in this regime,  we examine surface fluctuations on  circular droplets. 

As can be seen in Fig.~\ref{fig:etao}a-b and Supplementary Movie 10, the edge flow penetrates deeper into the chiral fluid as friction is reduced. The  dispersion relations for high and low friction droplets display the same trend, although the range of accessible wave vectors normalized by the penetration length ($k\delta$) is larger in the low friction case. An extension of our theory to circular geometries (see Supplementary Information) again accurately captures the dispersion relations for high friction (Fig.~\ref{fig:etao}a) and low friction (Fig.~\ref{fig:etao}b). 

The remarkable agreement between experiment and theory is however challenged when investigating the damping dynamics of the chiral waves. Experimentally, the damping rate $\alpha$ of chiral waves of wave vector $k$ is given by fitting a Lorentzian to the width of the power spectrum (see Supplementary Information); the resulting damping rates are shown in Fig.~\ref{fig:etao}c-d.
Our hydrodynamic theory predicts this damping rate to be proportional to surface tension. This is natural since surface tension flattens interfacial deformation: in the absence of inertia, the relaxation does not overshoot and capillary waves are overdamped. In the long wavelength limit ($k\delta\ll 1$), the damping rate $\alpha\sim (\gamma/\Gamma_u) |k|^3$ stems from the competition between surface tension and substrate friction. As seen in Fig.~\ref{fig:etao}c, in the high friction case we again find excellent agreement between theory and experiment, which provides a direct measurement of surface tension.  
The value we find, $\gamma = 2.3 \pm 0.2 \times 10^{-13}\rm\ N$, is consistent with an estimate based on magnetic interactions between rotating dipoles (see Supplementary Information).

In the case of low surface friction, however,  we observe a distinct new feature in the dissipation rate: a leveling off of the dissipation rate at short wavelengths which cannot be accounted for by the hydrodynamic theory discussed thus far, suggesting the presence of an additional  mechanism for surface wave dissipation in our chiral fluid.  
Seeking a hydrodynamic description, we recall that isotropic chiral fluids can in principle possess an additional stress in their constitutive relation, known interchangeably as  ``anomalous viscosity", ``odd viscosity" or ``Hall viscosity"~\cite{avron_viscosity_1995,avron_odd_1998,read_non-abelian_2009}. 
This non-dissipative, transverse stress is linked by Onsager relations to the breaking of time-reversal symmetry.

Theoretically, odd viscosity has indeed been shown to arise in the hydrodynamics of plasmas and systems of spinning molecules, `gears', as well as quantum Hall fluids and vortex fluids~\cite{radin_lorentz_1972,robinson_variational_1962,pitaevskii_physical_1981,wiegmann_anomalous_2014,banerjee_odd_2017}. We therefore conjecture our chiral fluid to support an additional Hall stress  $\sigma^{\rm o}_{ij}=\eta_o \left( \partial_i \epsilon_{jk}v_k + \epsilon_{ik}\partial_k v_j\right)$. 
In incompressible fluids such as the one considered here, the effect of odd viscosity can solely be seen at the edge. 
This is because in the bulk flow Hall stress is merely absorbed into the fluid pressure. 
The signature of odd viscosity in our chiral fluid is thus an additional boundary stress. The  component normal to the interface $\sigma_{nn}$ is given by 
\begin{equation}
\sigma_{nn}=\eta_{\rm o}\left(\partial_sv_{n}+\frac{v_s}{R(s)}\right),
\label{Eq:Hallstress}
\end{equation}
where $v_{n}$ (resp. $v_s$) is the velocity normal (resp. tangential) to the surface (see Fig.~\ref{fig:etao}e), and $R(s)$ is the local radius of curvature. 

In our system, where odd stress powers a boundary-layer edge flow, we thus expect odd viscosity to flatten surface deformation in a manner akin to surface tension, $\sigma^{\rm o}\sim \eta_{\rm o}v_s/R$. 
The excellent agreement between  our measurements and  predictions from a full hydrodynamic theory confirms this simplified picture and establishes  the presence of Hall viscosity in our colloidal chiral fluid (see Fig.~\ref{fig:etao}d, f-g). From the fit we obtain  $\eta_{\rm o}=1.4\pm 0.1 \times 10^{-8}\ \rm Pa\ m\ s$.

The clearly visible decrease in slope in the damping relation is the most visible signature of Hall viscosity in our data and can be understood on dimensional grounds. 
In the long wavelength limit, the wave relaxation time is controlled by the competition of either surface tension or Hall stress with substrate friction. Dimensionally this implies a scaling $\alpha \sim \vert k\vert^3$ since the ratios $\gamma/\Gamma_u$ and $\eta_{\rm o} v_s/\Gamma_u$ have dimension of volume per unit time. 
In contrast, in the short wavelength limit,  surface friction plays no role and damping stems from the competition of surface tension or Hall stress and bulk viscosities alone. In this case dimensional analysis requires linear scaling with wavenumber in the case of surface tension, and wave-number independence in the case of Hall stress (see Supplementary Information). This change in wavenumber dependence brings about a visible rollover to a decreased slope in the wave damping rate. 

We note that for small ranges of $k\delta \sim [-1,1]$, characteristic of spectral measurements in the presence of high surface friction, the leveling off cannot be seen and the relative roles of Hall viscosity and surface tension become hard to separate. This is the case for the damping shown in Fig.~\ref{fig:etao}c which can be fit well by both a non-zero and zero value of Hall viscosity (see Supplementary Information).

Having established the presence of Hall viscosity by examining wave damping, it follows to ask whether it has an  effect on wave propagation. The first term in Eq.~\eqref{Eq:Hallstress} suggests that Hall viscosity and surface tension could act together to support wave propagation. 
Surface tension acts on a sinusoidal surface deformation by pulling down peaks and pushing up troughs, generating an in-phase normal velocity component. 
The normal Hall stress $\partial_s v_n$ would then act out of phase on the inflection points of the  sinusoidal perturbation to propagate it in a chiral fashion. 
Our full theory confirms that this additional wave-pumping mechanism indeed exists and generates waves even in the absence of edge currents. 
However, for our hydrodynamic parameters, their effect on the dispersion is minimal. 

\section{An odd instability}

\begin{figure*}
\centering
\includegraphics[width=\textwidth]{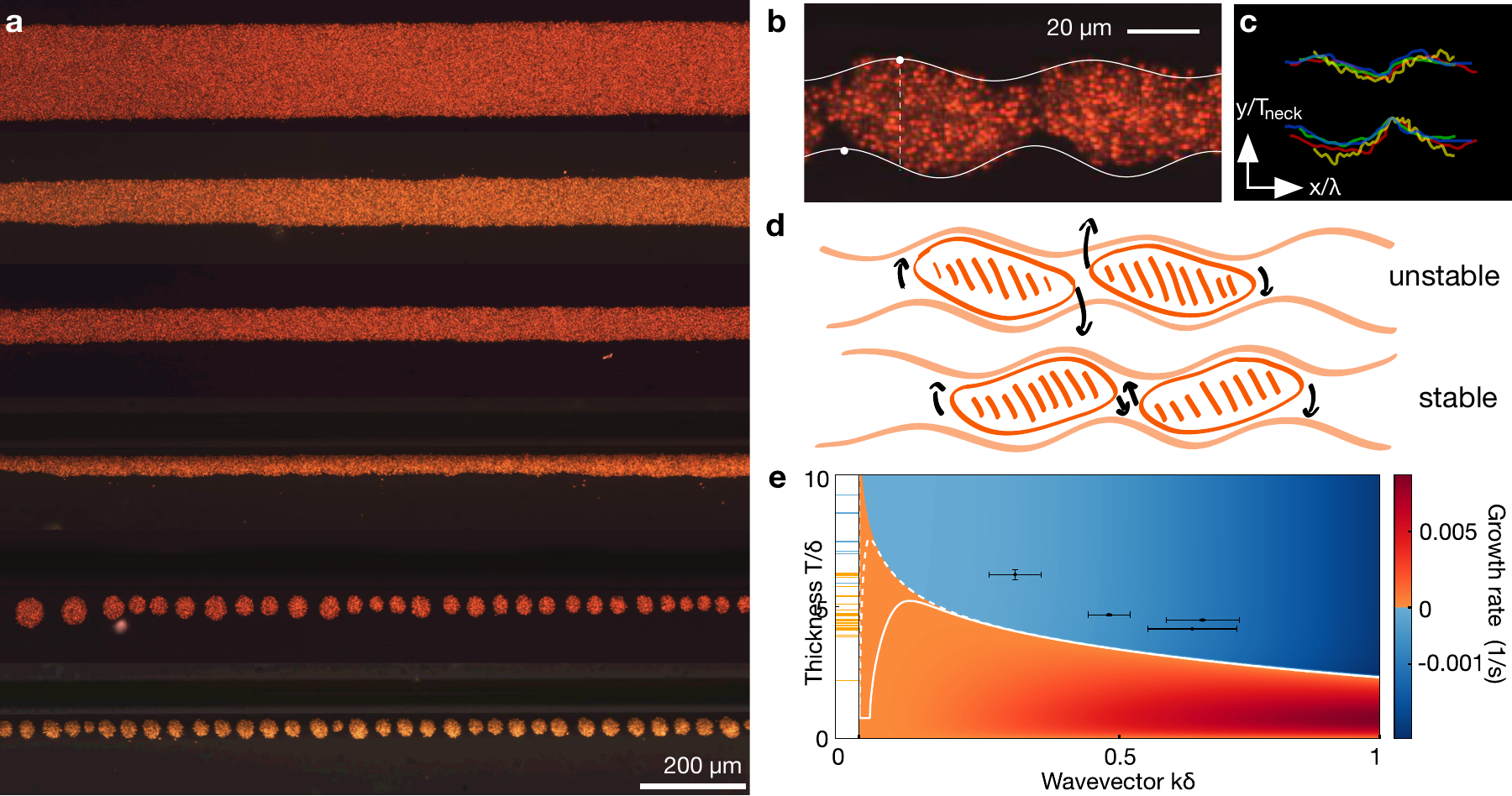}
\caption{{\bf A hydrodynamic instability.} 
{\bf a,} Strips of chiral fluid with different thicknesses. Above $32\, \rm \mu m$, the strips are stable, as observed over the course of $\ge 10$ minutes. Below $32\, \rm \mu m$ the strips break into droplets within 1 minute. 
{\bf b,} Chiral fluid strip approaching instability. Continuous white lines represent the sum of the most prominent Fourier modes of the strip outline. Relative phase difference between interfaces is emphasized by the two white dots and vertical dashed line.
{\bf c,} Overlay of strip outlines at four breakup points; each color corresponds to a different instability occurrence. $x$-axis is rescaled by the most prominent wavelength, $\lambda$. $y$-axis is rescaled by the thickness at the narrowest point, $T_{\rm neck}$. The relative phase between the top and bottom interface is consistent with theory.
{\bf d,} Schematic of the instability mechanism. Thin strips of chiral fluids are like a collection of elongated droplets rotating in the direction of the edge current. This leads to the breakup (top) or stabilization (bottom) of the strip. 
{\bf e,} Stability diagram, calculated with linear stability analysis using our experimentally extracted values for the hydrodynamic coefficients, with $\eta_{\rm o} = 0$ (see Supplementary Information for details). The thinner the strip, the larger the range of unstable wavelengths. A surface fluctuation at an unstable wavelength will grow exponentially: orange denotes a positive growth rate and blue denotes a negative growth rate, namely damping. Contour lines mark growth rates corresponding to 10 minutes (continuous) and 1 day (dashed). Black points represent experimental data from unstable strips; wavelengths were measured by Fourier-transforming the strip outline. Error bars in thickness correspond to the standard deviation in the measurement at various points. Error bars in wavelength correspond to the half-width of the Fourier peaks. Horizontal lines on $y$-axis mark the recorded strip thicknesses: orange and blue lines correspond to unstable and stable strips, respectively.
}
\label{fig:instability}
\end{figure*}

In much of the phenomenology we have discussed, surface dynamics are essentially boundary layer dynamics. 
Another natural question, then, is what happens when two boundary layers meet?
Draining fluid past a curved obstacle brings about the progressive thinning of a curved strip of chiral fluid, as shown in Fig.~\ref{fig:chiralfluid}g and Supplementary Movie 6.
The flow is smooth until the strip thickness becomes comparable to the penetration depth $\delta$; at that point the flow goes unstable, resulting in the formation of circular droplets. We study this novel pearling mechanism in experiment by creating a sequence of strips of decreasing thickness, as shown in Fig.~\ref{fig:instability}a
and Supplementary Movie 11. 
We find that over a period of 10 minutes the strips of chiral fluid are stable for thicknesses above $\sim 32\,\rm \mu m$ and unstable below.

Although visually reminiscent of the Rayleigh-Plateau instability of a thin fluid cylinder jet~\cite{eggers_nonlinear_1997}, this instability is fundamentally different. 
In our two-dimensional system, surface tension is a purely stabilizing force, as seen in the wave analysis discussed above.
Instead, the instability originates from the chiral surface dynamics of our  fluid. 
A visual signature of this origin is the consistent offset in the phase between top and bottom perturbations at the moment the instability occurs in all strips: Fig.~\ref{fig:instability}b shows one such example.

A linear stability analysis of a thin strip of chiral fluid quantitatively predicts the existence of unstable modes which 
consist of wave-like perturbations on the top and bottom surfaces that have a relative phase offset, as sketched in Fig.~\ref{fig:instability}d.   
These are accompanied by a stable mode with an opposite relative phase. 
The associated stability diagram is shown in Fig.~\ref{fig:instability}e, together with  our experimental observations. 
As the Hall stress has little effect on the stability of modes for small $\delta$ (see Supplementary Information), here we set  $\eta_o=0$.

An intuitive picture for the mechanism driving the instability is illustrated in Fig.~\ref{fig:instability}d.
The geometry of a thin slab with out-of-phase perturbations on the top and bottom surfaces can be approximated by a collection of elongated droplets of chiral fluid all canted in the same direction. 
Droplets of this kind rotate in the direction of the edge current, in this case clockwise (see Fig.~\ref{fig:chiralfluid}d and Supplementary Movie 3). Depending on the phase difference between the two interfaces, the rotation of these effective droplets will either increase the amplitude of the perturbation, resulting in the breakup of the strip (top); or decrease the amplitude of the perturbation and restore the flat interface (bottom). 
The consistent observation of this phase relation between the top and bottom perturbations across many experiments of strips going unstable (Fig.~\ref{fig:instability}c) further corroborates  our theoretical picture of the instability.

We have broken parity symmetry at the microscopic level in a colloidal chiral fluid, resulting in the emergence of an odd stress that in turn generates lively surface flows.
Likewise, we have broken time reversal symmetry, giving rise to Hall viscosity, a dissipationless transport property which has thus far remained experimentally elusive. 
The combination of these features drives rich interfacial dynamics with no analogues in conventional fluids.
These dynamics include the uni-directional propagation and anomalous attenuation of surface waves and an asymmetric pearling instability. 
In principle, these chiral phenomena can be tuned, for instance by altering the colloidal particles' shape and their effective interactions.
Beyond enabling the study of universal aspects of a new class of hydrodynamics, colloidal chiral fluids provide a platform for engineering active materials with so far untapped, `odd' behaviors~\cite{avron_odd_1998,avron_viscosity_1995,banerjee_odd_2017}.

\bibliographystyle{naturemag_nourl}


\end{document}